\documentclass[twocolumn,prl,aps]{revtex4}
\usepackage{braket,xspace,amsmath,amsfonts,amsthm,amssymb,amsbsy,graphics,color,graphicx,bbm}
\usepackage{epigraph}
\usepackage{soul} 
\setstcolor{red}

\usepackage[breaklinks=true]{hyperref}
\hypersetup{
  colorlinks   = true, 
  urlcolor     = blue, 
  linkcolor    = blue,
  citecolor   = blue 
}
\usepackage{framed}

\begin{document}
%
%
%
%
%


Kedem's Comment \cite{k} on our Letter \cite{FerCom13} boils down to the claim that because we did not consider imaginary or complex weak values, our conclusions do not hold generally.  The first example of analysis of complex weak values, along the same lines as we have provided in our Letter, was done by Knee and Gauger \cite{KneGau13}.  It was shown that the same kind of inequality that Tanaka and Yamamoto \cite{TanYam13} and we \cite{FerCom13} have derived still holds.  In a follow up paper \cite{ComFerJia13}, we proved that for any quantum operation post-selection cannot improve estimation. The analysis there, which is a straightforward generalization of that we provided in \cite{FerCom13}, includes the most general evolution allowed by quantum theory---a quantum operation. Thus, it includes all possible experimental arrangements and consequently includes any defined quantities such as real weak values, imaginary weak value even quaternionic or topological weak values, which, in our view, only seems to obscure what is a simple problem of statistics.

Vaidman's Comment \cite{v} can be deconstructed in to two distinct logical fallacies.  Below we state the logical fallacies, summarize them and then comment on them.

\noindent {\bf Part 1 (the straw man fallacy):} \emph{Vaidman cherry picks a quote from our Letter stating that ``post-selection cannot aid in detect and estimate'' and then goes on to quote experiments which claim the opposite.}

Vaidman has a point that our wording ``cannot aid'' might suggest to some that have not read and understood our Letter that the experiments he cites are not possible, whereas clearly they are possible since they were performed.  However, Vaidman is attacking a straw man as this was the wrong interpretation of our comment.  Our intended meaning was that post-selection cannot aid \emph{above and beyond} the usual quantum limits for estimation.  In fact, we prove that post-selection and arranging for a large weak value actually decrease estimation and detection accuracy.
\noindent{\bf Part 2 (the irrelevant conclusion):} \emph{Vaidman goes on to say that the reason our conclusion is questionable is that in optical implementations of weak value amplification the total number of photons is not the relevant number because of detector saturation.}

We will argue that detector saturation is a red herring.  Vaidman claims that because in his particular experimental setup \cite{Kedem} the detector saturates at a power level less than the laser, he only needs to count as a resource those photons that are detected.  This sentiment is equivalent to the conclusion of Ref.~\cite{zhu} which Vaidman quotes ``the probability reduced by post-selection need not be considered....''

Of course, we are not the resource counting police and one is free to count whatever resources one wishes; however, with post-selection, one can always ignore resources which make one look bad.  If we count all photons entering the experiment, then WVA is not optimal---this is our result \cite{FerCom13}.  However, if we ignore the discarded data and only count as a resource the post-selected data, then WVA appears to improve matters.  One might also note that if we count only the money spent on winning lottery tickets, our net income looks pretty good---the bank disagrees, however.

There is also something more broadly interesting at play here.  In theoretical analyses, one must necessarily ignore at least some experimental contingencies.  Then, when the optimal experiment is determined and implemented, various work-arounds can be employed to circumvent these contingencies.  However, and this is the key point, the \emph{reason} that one experiment outperforms another is due to physics, not the particular way in which some technical limitation was overcome and certainly not the way the data was processed.

 In a follow up paper \cite{ComFerJia13}, we formulated the following desiderata for rigorous analysis of probabilistic metrology (a generalization which includes WVA):
\begin{enumerate}
\item[D1:] Choose a performance metric, and apply it uniformly to all data.
\item[D2:] Include the success probability correctly in the analysis.  This should happen automatically if the problem is set up properly.  
\item[D3:] Compare the performance of probabilistic protocols with deterministic protocols, using the same performance metric for all cases.  If possible, compare with the optimal deterministic protocol, which sets a quantum limit on estimation as measured by the chosen performance metric.
\end{enumerate}
 If an analysis that adheres to these desiderata is to conclude that probabilistic metrology improves parameter estimation it would require violating one of the following two principles:
 \begin{enumerate}
\item[P1:] A suboptimal strategy cannot achieve optimal performance;
\item[P2:] Information cannot be increased by throwing some of it away.
\end{enumerate}
There is little reason to believe that the presence of detector saturation, when analyzed properly, will invalidate these principles.  The onus is on the probabilistic metrology and WVA community to detail how such obvious principles can be violated.  Results without the context that comes from adhering to the desiderata have little call on the attention of the quantum metrology community.

\vspace{0.5pc}
\begin{acknowledgements}
We acknowledge financial support from NSF Grant Nos.~PHY-1212445 and PHY-1005540, by ONR Grant No.~N00014-11-1-0082.  CF acknowledges funding from NSERC.
\end{acknowledgements} 

\vspace{1pc}

\noindent Christopher Ferrie and Joshua Combes

\vspace{0.5pc}
\noindent Center for Quantum Information and Control, University of New Mexico, Albuquerque, New Mexico, 87131-0001


\begin{thebibliography}{99}





\bibitem{k}
Y. Kedem, \href{http://arxiv.org/abs/1402.1352}{arXiv:1402.1352}.


\bibitem{FerCom13}
C. Ferrie and J. Combes, \href{http://link.aps.org/doi/10.1103/PhysRevLett.112.040406}{Phys. Rev. Lett. {\bf 112}, 040406 (2014)}.

\bibitem{Kedem}
X-Y Xu, Y. Kedem, K. Sun, L. Vaidman, C.-F. Li, and G.-C. Guo, \href{http://dx.doi.org/10.1103/10.1103/PhysRevLett.111.033604}{Phys. Rev. Lett. {\bf 111}, 033604 (2013)}. 


\bibitem{KneGau13}
G. C. Knee and E. M. Gauger, \href{http://arxiv.org/abs/1306.6321}{arXiv:1306.6321}.

\bibitem{TanYam13}
S. Tanaka and N. Yamamoto, \href{http://dx.doi.org/10.1103/PhysRevA.88.042116}{Phys. Rev. A {\bf 88}, 042116 (2013)}.


\bibitem{ComFerJia13}
J. Combes, C.~Ferrie, Z. Jiang, C. M. Caves, \href{http://arxiv.org/abs/1309.6620}{arXiv:1309.6620}.

\bibitem{v}
L. Vaidman, \href{http://arxiv.org/abs/1402.0199}{arXiv:1402.0199}.


\bibitem{zhu}
X. Zhu, Y. Zhang, S. Pang, C. Qiao, Q. Liu, and S. Wu,
\href{http://pra.aps.org/abstract/PRA/v84/i5/e052111}{Phys. Rev. A 84, 052111 (2011)}.

 
\end{thebibliography}
\end{document}